\documentclass{ieeeaccess}
\usepackage{cite}
\usepackage{amsmath,amssymb,amsfonts}
\usepackage{algorithmic}
\usepackage{graphicx}
\usepackage{xcolor}
\usepackage{textcomp}
\usepackage{url}
\usepackage{makecell}
\usepackage{changepage}
\def\BibTeX{{\rm B\kern-.05em{\sc i\kern-.025em b}\kern-.08em
    T\kern-.1667em\lower.7ex\hbox{E}\kern-.125emX}}
\begin{document}
\history{Date of publication xxxx 00, 0000, date of current version xxxx 00, 0000.}
\doi{10.1109/ACCESS.2023.0322000}

\title{Forensic Camera Identification: Effects of Off-Nominal Exposures}
\author{\uppercase{Abby Martin}\authorrefmark{1},
\uppercase{Roy Maxion}\authorrefmark{2}, \IEEEmembership{IEEE Fellow}, and \uppercase{Jennifer Newman}\authorrefmark{3},
\IEEEmembership{IEEE Senior Member}}

\address[1]{Department of Mathematics, Iowa State University, Ames, IA 50011 USA (e-mail: abby2@iastate.edu)}
\address[2]{Department of Computer Science, Carnegie Mellon University, Pittsburgh,
PA 15213 USA (e-mail: maxion@cs.cmu.edu)}
\address[3]{Department of Mathematics, Iowa State University, Ames, IA 50011 USA (email: jlnewman@iastate.edu)}
\tfootnote{This work was funded (or partially funded) by the Center for Statistics and Applications in Forensic Evidence (CSAFE) through Cooperative Agreements 70NANB15H176 and 70NANB20H019 between NIST and Iowa State University, which includes activities carried out at Carnegie Mellon University, Duke University, University of California Irvine, University of Virginia, West Virginia University, University of Pennsylvania, Swarthmore College and University of Nebraska, Lincoln.
}

\markboth
{Martin \headeretal: Forensic Camera Identification: Effects of Off-Nominal Exposures}
{Author \headeretal: Forensic Camera Identification: Effects of Off-Nominal Exposures}

\corresp{Corresponding author: Abby Martin (e-mail: abby2@iastate.edu).}

\begin{abstract}
\label{Abstract}

Photo-response non-uniformity (PRNU) is a technology that can match a digital photograph to the camera that took it. Due to its use in forensic investigations and use by forensic experts in court, it is important that error rates for this technology are reliable for a wide range of evidence image types. In particular, images with off-nominal exposures are not uncommon. This paper presents a preliminary investigation of the impact that images with different exposure types — too dark or too light — have on error rates for PRNU source camera identification. We construct a new dataset comprised of 8400 carefully collected images ranging from under-exposed (too dark) to nominally exposed to over-exposed (too bright). We first establish baseline error rates using only nominally exposed images, resulting in a true-positive rate of 100\% and a true-negative rate of 99.92\%. When off-nominal images are tested, we find striking results: the true-negative rate for under-exposed images is 99.46\% (a false-positive rate of roughly one in two hundred, typically unacceptable in a forensic context), and for over-exposed images the true-positive rate falls to 82.90\%. Our results highlight the importance of continued study of error rates for the PRNU source camera identification to assure adherence to the high standards set for admissibility of forensic evidence in court.

\end{abstract}

\begin{keywords}
Camera Identification, Digital Forensics, Image Forensics, PRNU, Questioned Images
\end{keywords}

\titlepgskip=-21pt

\maketitle

\section{Introduction}
\label{Introduction}
\PARstart{C}{onsider} a photo found at a crime scene. The photo can be traced to the camera that took it using a phenomenon known as photo response non-uniformity (PRNU). PRNU is a unique and persistent pattern (fingerprint) of variabilities of voltage levels across the pixels in a digital camera’s photosensitive image sensor, and this pattern appears in each photo the camera takes. The PRNU fingerprint of the camera’s sensor is compared with the PRNU fingerprint of a photo, producing a score measured against an accepted fixed threshold to determine a match (or non-match) between the camera sensor and the photo. 

The PRNU source-camera-identification algorithm is based on a large-scale study of images from Flickr \cite{GoljanFridrichFiller2009}, and is considered accurate across a variety of images. Flickr is a public website for users to share images. Due to self-curation (users uploading their best images), Flickr has a negligible proportion of off-nominal exposure images \cite{IulianiFontaniPiva2021} (e.g., over- or under-exposed, like the examples in Fig. \ref{ExposureSceneExamples} \cite{NewmanLinChenReindersWangWuGuan2019}). Thus, results from Flickr data may have overlooked limitations, including generalizing error rates to data not represented in the Flickr dataset (e.g., off-nominal exposure types).

The exclusion of off-nominal data when testing forensics tools has real-world consequences. Studies performed by the forensic community play a critical role supporting admissibility of expert witness testimony under federal law (including estimating error rates), as introduced by Daubert v. Merrell Dow Pharmaceuticals \cite{DaubertvMerrell}. 
A large-scale study \cite{GoljanFridrichFiller2009} established the court-approved PRNU source camera identification algorithm using Flickr images to determine error rates. However, many forensic tools have opportunities for improvement due to a host of factors such as lack of standardize corpora \cite{GarfinkelFarrellRoussevDinolt2009}, non-existent tool validation procedures \cite{HughesKarabiyik2020}, etc. We were unable to find any studies that determined error rates for off-nominal exposure images for the PRNU source camera identification algorithm. Additionally, the data in \cite{GoljanFridrichFiller2009} is no longer available \cite{Goljan2023}. This motivated our investigation of the PRNU source-camera-identification algorithm applied to known off-nominal image data. 

Our contribution is a methodology to evaluate the response of the PRNU source-camera-identification algorithm to off-nominal exposure images. Forensic experts representing both the prosecution and defense can use publications incorporating this methodology to better represent the error rates associated with specific evidence images, especially the false-positive error rate (incorrectly matching an image with a camera). Our methodology can be adapted to estimate PRNU source-camera-identification error rates for other types of off-nominal images as well. This process can be integrated into the estimation of error rates for a tool or technique to satisfy the Daubert criteria.

The rest of the paper is organized as follows: Section \ref{ProblemAndApproach} outlines the research problem and approach; Section \ref{Terminology} clarifies language used throughout the remainder of the paper; Section \ref{BackgroundAndRelatedWork} is a discussion of the forensic development of PRNU source camera identification; Section \ref{OverviewOfPRNUCameraIdentification} provides a detailed overview of the PRNU source-camera-identification algorithm used in this paper; Section \ref{Data} describes the data and characterizes under- and over-exposed images; Section \ref{Methods} outlines the methodology of the experiments performed; and Section \ref{Results} presents the results. In Section \ref{Limitations}, we share the limitations of the experiments, followed by a discussion of implications of our results in Section \ref{Discussion}. We conclude with a summary in Section \ref{Conclusions}.


\section{Problem and Approach}
\label{ProblemAndApproach}

Recognizing that prior work did not account for off-nominal images (e.g., too light or too dark), the present work aims to determine whether off-nominal images skew the true positive and true negative rates previously reported in \cite{GoljanFridrichFiller2009}. We implement the PRNU algorithm presented in that work to establish error rates for off-nominal images. In Section \ref{Methods}, we present a rigorous evaluation of the error rates for this source-camera-identification algorithm when the image exposures are off-nominal, including a sensitivity analysis to determine how proportions of off-nominal images in the dataset can impact error rate estimates. In principle, this framework could also be applied to other types of off-nominal image data (e.g., digital zoom or out-of-focus).


\section{Terminology}
\label{Terminology}

The image of unknown origin, which could be considered evidence in a forensic case, is referred to as the \textit{questioned image}. The camera believed to have taken the questioned image is referred to as the \textit{specific camera}. To avoid repetitious language, when we refer to an \textit{image}, we mean the questioned image (unless otherwise specified). Similarly, when we refer to a \textit{camera}, we mean the specific camera.

We focus on the {\it source-camera-identification problem}, which aims to identify the specific camera that captured a questioned image. We do not address the {\it camera-model-identification problem}, which attempts to identify only the model of the camera used to collect the questioned image. As an example, source camera identification could conclude an image came from a specific iPhone6s camera, whereas camera-model identification would only be able to conclude the image was from some iPhone6s camera (i.e., \emph{this} camera versus this \emph{kind} of camera). For simplicity, when we refer to {\it camera identification} we are referring to the source-camera-identification problem. We use {\it PRNU cameraID algorithm} to refer specifically to the source-camera-identification algorithm as established in the large-scale study \cite{GoljanFridrichFiller2009}, which is the algorithm used throughout this paper.

\section{Background and Related Work}
\label{BackgroundAndRelatedWork}

Photo response non-uniformity (PRNU) is a persistent artifact in digital images due to imperfections in the camera-sensor manufacturing process \cite{LukasFridrichGoljan2006}. When light impinges on the photosensitive portion of a pixel, called the photodiode, the pixel responds by generating a current in proportion to the number of photons striking it. However, the imperfections result in consistent differences from the mean values of currents among the pixels, and it is this pattern of responses – the PRNU – that is unique to each camera sensor. This pattern remains constant and present in each image; therefore, PRNU can be considered part of the sensor’s \textit{fixed pattern noise}.

The first digital image sensor, a charged-coupled device (CCD), was invented in 1970 \cite{BoyleSmith1970}, and within a year large variabilities in dark signal (thermal noise when no light is falling on the sensor) were observed \cite{TompsettEtAl1971}. The first use of fixed pattern noise to identify an individual digital camera sensor appeared in \cite{KurosawaKurokiSaitoh1999} using the dark signal. This method assumes scenes that are very dark, so it is not useful for images taken under typical lighting conditions.

In 2005, the first computational method was introduced for extracting the PRNU from an image for the purpose of source camera identification \cite{LukasFridrichGoljan2005}. Since the PRNU can be extracted from photos taken in typical lighting environments, researchers shifted to the PRNU as a camera fingerprint. Several improvements followed, including: introducing a maximum likelihood estimator to improve the camera fingerprint calculation \cite{ChenFridrichGoljan2007}; changing to the peak-to-correlation energy (PCE) ratio as a similarity metric \cite{GoljanFridrich2008}; and discovering that (unlike correlation scores) a threshold value based on the PCE does not need to change for each camera fingerprint estimation \cite{Goljan2008}.

PRNU-based camera identification became standardized for court use in a large-scale study performed on 1,053,580 JPEG images from Flickr representing approximately 6,896 cameras over 150 models \cite{GoljanFridrichFiller2009}. This work used the distribution of PCE scores between the questioned images and the camera fingerprints to set a threshold based on the false positive rate (FPR) to ensure acceptable true negative rates (TNR) and true positive rates (TPR), resulting in a recommended PCE threshold of 60. The overall identification rates given in \cite{GoljanFridrichFiller2009} are a TPR of 97.62\% and a TNR of 99.9976\%. Our goal is to use the standard algorithm set by this work to establish error rates for a very different set of data. In Section \ref{OverviewOfPRNUCameraIdentification}, we give details of the PRNU source-camera-identification (PRNU cameraID) algorithm as established in \cite{GoljanFridrichFiller2009}.

Research since 2009 has continued to modify and question PRNU camera identification. Use of the PRNU has expanded to fields such as forensic countermeasures for forged PRNU information \cite{GoljanFridrichChen2010, GoljanFridrichChen2011}, image anonymity techniques \cite{IrshadLawLooHaider2023}, convolutional neural networks (CNN)-based forgery detection \cite{CozzolinoVerdoliva2020}, and user authentication \cite{MaierErbMullanHaupert2020}. Many papers have proposed changes to the PRNU cameraID algorithm: enhancements to fingerprint estimation \cite{CortianaConotterBoatoDeNatale2011, LiLiuMaWangQinWuLi2023, FernandezMenduinaPerezGonzalezMasciopinto2023, MehrishSubramanyamVenkataSabu2018}; different similarity measures \cite{ReindersLinChenGuanNewman2020}; and use of machine-learning methods such as CNNs \cite{FanfaniPivaColombo2022}. Other work has focused on the impact of various image artifacts on PRNU camera identification, including: vignetting \cite{LiSatta2012}; JPEG compression \cite{GoljanChenComesanaFridrich2016}; proprietary image processing \cite{BaracchiIulianiNenciniPiva2020, MontibellerPerezGonzalez2023, AlbisaniIulianiPiva2021, JoshiKorusKhannaMemon2020}; color saturation \cite{Henry2016}; as well as gamma correction, contrast enhancement, histogram equalization, and white balance \cite{SamarasMygdalisPitas2016}. One large-scale study of over 33,000 images from Flickr found several devices with low true negative rates (e.g., 0.8\% for the Fujifilm X-T30 camera) \cite{IulianiFontaniPiva2021} for images from the same camera model as the specific camera.

Concerns about PRNU-based camera identification have been raised in preliminary investigations of exposure settings, specifically ISO. A brief study of the effects of ISO values on PRNU camera identification found that ISO impacts noise levels, gray levels, and correlation values \cite{LinEtAll2018}. The Warwick database \cite{QuanLiZhouLi2020}, which contains images with a variety of ISO values, establishes that forgery detection and correlation-predictor values are impacted by ISO \cite{QuanLi2021}. However, the Warwick investigation of ISO relies on correlation predictors to identify forged regions of images and does not implement a decision threshold \cite{QuanLi2021}. The standard PRNU cameraID algorithm in \cite{GoljanFridrichFiller2009} relies on PCE scores and a threshold of 60, not correlation predictors. Additionally, the authors in \cite{QuanLi2021} vary both the ISO and exposure time settings to ensure similar exposure values between images of the same scene, meaning all images have a consistent visual brightness. The research question in \cite{QuanLi2021} differs from ours: their paper asks whether correlation predictors and forgery detection are impacted by ISO (the exposure type of the image remains constant by changing the exposure time to compensate for the changes in ISO). We ask whether the PCE score and camera identification are impacted by changing the exposure type (i.e., auto-, under-, or over-exposed images). For our data, we intentionally vary the ISO and exposure time to gather brighter and darker versions of the same scene, thus varying the exposure value and brightness of the image (see Fig. \ref{ExposureSceneExamples}). 

In order to establish error rates for off-nominal images, we must adhere to the algorithm used by forensic experts. To the best of our knowledge, this is the algorithm implemented in the large-scale study \cite{GoljanFridrichFiller2009}. Whereas these recent works \cite{LinEtAll2018, QuanLi2021} \cite{QuanLiZhouLi2020} have addressed the impact of ISO on correlation, forgery detection, and correlation predictor values, we do not know of any work that has addressed the problem of over- or under-exposed images and their effect on error rates for the PRNU cameraID algorithm. This paper tackles exactly that problem.


\section{Overview of PRNU Camera Identification}
\label{OverviewOfPRNUCameraIdentification}

We refer to the protocol followed in \cite{GoljanFridrichFiller2009} as the PRNU cameraID algorithm. Generally, this algorithm consists of three parts: (1) estimating the PRNU fingerprints of the camera and of the image; (2) calculating the peak-to-correlation energy (PCE) ratio between these two fingerprints; (3) comparing the PCE score with a threshold of 60 to determine whether (or not) the camera captured the image. The details of this process are described in the remainder of this section.

\subsection{Estimate Fingerprints Using PRNU Noise Model}
\label{EstimateFingerprintsUsingPRNUNoiseModel}
This section gives a summary of the fingerprint extraction algorithms as presented in \cite{GoljanChenFridrich2007}. To compare a camera with an image, we must estimate the PRNU noise from both the camera sensor and the image. Let $I$ be an image, and let $I_0$ be the corresponding “noiseless image” which would result from a sensor without any imperfections. Similarly, let $K$ be the true PRNU noise component of the camera. Note that all multiplication in this section is performed element-wise, thus the image $I$ is modeled:

\begin{equation}
\label{Eq1}
    I = I_0 + I_0K + \Theta,
\end{equation} 

where all noise components besides the PRNU are denoted by $\Theta$ \cite{GoljanChenFridrich2007}.

However, it is not feasible to obtain the noiseless image $I_0$ or the true PRNU fingerprint $K$. In order to approximate the PRNU component of the image noise, a high-pass filter is applied to suppress scene content and reduce non-unique low-frequency patterns included in the noise fingerprint of an image, such as intensity gradient and vignetting. The Daubechies wavelet denoising filter \cite{Daubechies1988} was originally chosen because it produced the best experimental results, likely due to its superior scene suppression (particularly for edges that appear within an image) \cite{LukasFridrichGoljan2006}. Hence, $F$ is the Daubechies denoising filter applied to the image.

The noise $W$ for image $I$ can be estimated \cite{ChenFridrichGoljan2007}:
\begin{equation}
\label{Eq2}
    W = I - F(I). 
\end{equation} 

This denoising step is followed by additional image processing of the grayscale image to remove non-unique artifacts (NUAs) due to JPEG-compression and camera-model fixed-pattern noise. NUAs can increase the similarity between images from different cameras and thus contribute to a lower TNR \cite{Fridrich2013}. The image fingerprint estimate is therefore calculated as:
\begin{equation}
\label{Eq2a}
    Q = G(W),
\end{equation} 
where $G$ represents these additional image processing operations.

The camera-fingerprint estimation follows a similar process, but is calculated using a set of several images. This improves the PRNU noise estimate in \eqref{Eq2} by implementing a maximum likelihood estimate of the true camera fingerprint $K$ using several images. First, we apply $F$, the Daubechies denoising filter, to all $N$ images. Let $I^{(i)}$ for $i \in \{1,2,...,N\}$ be the set of images used to estimate $\hat{K}$, the camera fingerprint. First, estimate the image fingerprint for each $I^{(i)}$ as before, $W^{(i)} = I^{(i)} - F(I^{(i)})$. Then the camera fingerprint $\hat{K}$ 
is estimated \cite{ChenFridrichGoljanLukas2008}:

\begin{equation}
\label{Eq3}
    \hat{K} = G\left(\frac{\sum_{i=1}^N W^{(i)}I^{(i)}}{\sum_{i=1}^N (I^{(i)})^2}\right),
\end{equation} 

where we used $N = 30$ images, and $G(\cdot)$ represents the additional processing performed after calculating the maximum likelihood estimate of the camera fingerprint to remove NUAs due to the camera model and JPEG compression, as done for the fingerprint estimate for a single image.

From a careful reading of the available code \cite{GoljanChenComeanaFridrichMATLAB} in MATLAB \cite{MATLAB2022}, we observe that saturated pixels of an image are excluded from the camera fingerprint estimate, where saturated pixels are characterized by the maximum intensity value of the image (at least 250) with at least one neighboring pixel of equal intensity. Although we intentionally collected very bright images, none of the pixels in our over-exposed (or any exposure type) image set meet these standards for a saturated pixel.

\subsection{Peak-To-Correlation Energy (PCE) Calculation}
\label{PCECalculation}

When we calculate the similarity between camera and image fingerprints, we use the \textit{signed} PCE score given in Equation (8) of \cite{GoljanChenComesanaFridrich2016} as well as in the MATLAB implementation \cite{GoljanChenComeanaFridrichMATLAB}. Using the signed PCE score differs slightly from the PCE calculation in \cite{GoljanFridrichFiller2009}, because if the peak correlation is negative, then the PCE score will be negative. Negative PCE scores could change the estimated probability density function (Equation (12) in \cite{GoljanFridrichFiller2009}) used to set the threshold of 60. An alternative distribution could impact the chosen PCE decision threshold, which would change the estimated error rates. Although it is unlikely that the signed PCE would impact the PRNU cameraID error rate estimates, it is worth noting this change from the algorithm as initially implemented \cite{GoljanFridrichFiller2009}.

\subsection{PRNU Source Camera Identification Algorithm Overview}
\label{PRNUCamIDOverview}

\begin{figure}[!h]

\includegraphics[scale = 1]{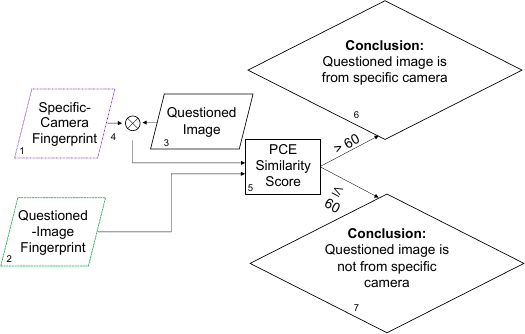}

\caption{\textbf{PRNU cameraID Algorithm.} Boxes 1 and 2 are the fingerprints estimated for the camera and questioned image, respectively. Operation 4 is an element-wise multiplication between two matrices representing the camera fingerprint (Box 1) and questioned image (Box 3). If the PCE (Box 5) is above 60, conclude that the image was taken by the specific-camera (Box 6); otherwise not (Box 7).}
\label{Standard PRNU Algorithm}
\end{figure}

To replicate the algorithm in \cite{GoljanFridrichFiller2009}, we used the MATLAB \cite{MATLAB2022} code provided by the same authors \cite{GoljanChenComeanaFridrichMATLAB}. An overview of the PRNU cameraID algorithm is shown in Fig. \ref{Standard PRNU Algorithm}. The inputs are two fingerprints, one from the camera (Box 1 in Fig. \ref{Standard PRNU Algorithm}) and one from an image (Box 2 in Fig. \ref{Standard PRNU Algorithm}), and the questioned image (Box 3 in Fig. \ref{Standard PRNU Algorithm}). Recall that the image fingerprint is estimated from a single image, and the camera fingerprint is estimated from 30 images. 

The PCE score is calculated between the image fingerprint and the product of the camera fingerprint and the image pixel intensities  (operation 4 on Boxes 1 and 3 in Fig. \ref{Standard PRNU Algorithm}). If the PCE score is greater than 60, conclude that the image came from the camera under test (Box 6 in Fig. \ref{Standard PRNU Algorithm}). Otherwise, conclude that the image originated elsewhere (Box 7 in Fig. \ref{Standard PRNU Algorithm}).

\section{Data}
\label{Data}

The dataset from \cite{GoljanFridrichFiller2009} is no longer available \cite{Goljan2023}, so we collected 8400 images from StegoAppDB \cite{NewmanLinChenReindersWangWuGuan2019} with specific ISO and exposure-time settings relative to the auto-exposure settings. This section provides our characterization of off-nominal exposure types, as well as the data sources, data protocol, and theoretical and experimental support for our labeling decision of over- and under-exposed images.

\subsection{Characterization of Over- and Under-Exposed Images}
\label{CharacterizationOEUE}

\begin{figure*}[h!]

\centerline{\includegraphics[scale = 1]{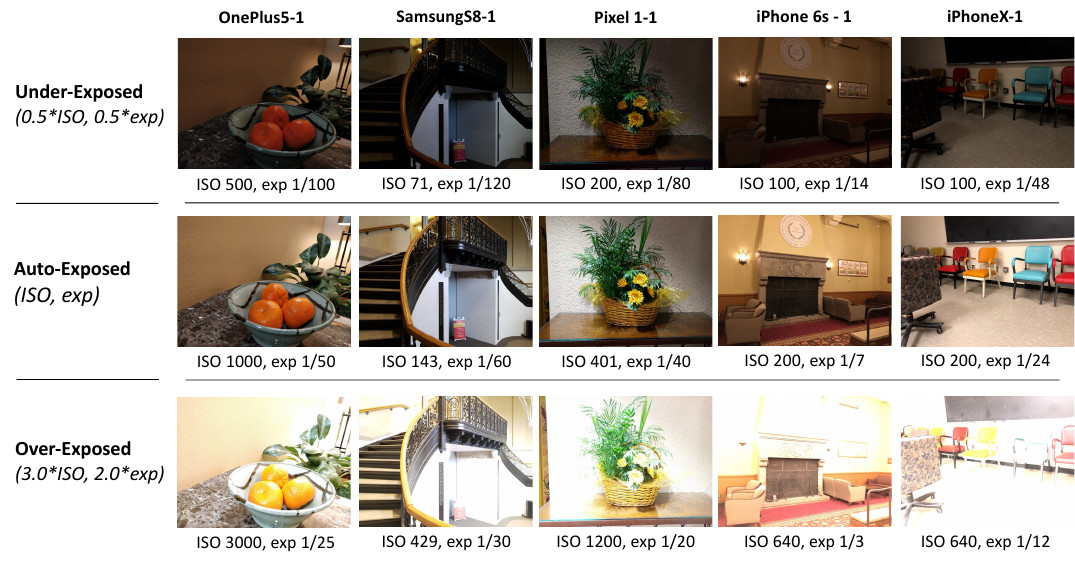}}
\caption{\textbf{Off-Nominal and Nominal Exposure Examples.} The first row is a sample of under-exposed images, the second row is a sample of auto-exposed images, and the third row is a sample of over-exposed images from our dataset \cite{NewmanLinChenReindersWangWuGuan2019}. The pair (ISO, EXP) denotes the auto-exposure ISO and exposure time settings, respectively. The programmed (ISO, EXP) values relative to the auto-exposed settings are given to the left of the images under the exposure type, and the recorded values for the actual data are provided underneath the photo itself. The model and device number that acquired the trio of images is given at the top of each column of images (model-device number). Five different devices provided the sample data.}
\label{ExposureSceneExamples}
\end{figure*}

A selection of over-, under-, and auto-exposed images from our dataset is shown in Fig. \ref{ExposureSceneExamples}, along with their ISO and exposure time settings. These are the three exposure types of data we use in our experiments. Perceived brightness of objects varies, so we characterize an off-nominally exposed image by comparing its level of brightness with its nominally exposed version. Specifically, we say an image is over-exposed (third row in Fig. \ref{ExposureSceneExamples}) if its overall visual appearance is noticeably brighter than its auto-exposed counterpart (second row in Fig. \ref{ExposureSceneExamples}). Similarly, we say an image is under-exposed (first row in Fig. \ref{ExposureSceneExamples}) if its overall visual appearance is noticeably darker than its auto-exposed counterpart. See Section \ref{Instrumentation} for the formulaic relation between auto- and off-nominal exposure settings. 

The intentional selection of exposure settings relative to auto-exposure is one supporting argument for an image being over- or under-exposed, but we provide two additional arguments that strengthen our characterization of exposure type: (1) the exposure value \cite{Lee2005} quantifies the brightness of each off-nominal image relative to its auto-exposed version; and (2) a visual inspection by three human judges assesses the agreement between the captured exposure settings and a human visual assessment of the brightness level. With these additional arguments, we obtain an acceptable level of certainty that the characterizations of exposure type for over-, under-, and auto-exposed images in StegoAppDB are indeed consistent and are satisfactory for testing the PRNU cameraID algorithm for nominal and off-nominal exposures.

\subsection{Apparatus \& Instrumentation}
\label{AppInst}

\subsubsection{Apparatus}
\label{Apparatus}
Twenty-eight smartphone cameras were used to acquire the images used in this research; they are cataloged in Table \ref{DataDevices}.

\subsubsection{Instrumentation}
\label{Instrumentation}
The controlled collection of images for hundreds of scenes using 28 separate smartphone cameras in StegoAppDB \cite{NewmanLinChenReindersWangWuGuan2019} requires some amount of automation, not least because the images were taken with specific ISOs and exposure times. For this reason, an application (app) called Cameraw (pronounced camer-raw) was written in Apple’s Swift language \cite{Swift2014} for Apple devices, and in Java \cite{Java2000} for Android devices.

\begin{table}[hbt!]
\small
 
\caption{Instruments, OS, protocols.
Parenthetical numbers indicate multiple instances of the camera model. 
100 over-, 100 auto-, and 100 under-exposed images (300 images total) per camera.}
\label{DataDevices}
 
\begin{center}
\begin{tabular}{|l|c|c|c|c|}
\hline
 
& \textbf{Camera (\# of Devices)} & \textbf{OS} & \textbf{Protocol}  \\ \hline
1 & Pixel1 (4)       & Android & A (Same Scene) \\ \hline
2 & Pixel2 (4)       & Android & A (Same Scene) \\ \hline
3 & iPhone6s (4)     & iOS     & A (Same Scene) \\ \hline
4 & iPhone7 (4)      & iOS     & A (Same Scene) \\ \hline
5 & iPhone8 (2)      & iOS     & A (Same Scene)  \\ \hline
6 & OnePlus5 (2)     & Android & B (Unique Scene)  \\ \hline
7 & SamsungS8 (2)   & Android & B (Unique Scene) \\ \hline
8 & iPhone6sPlus (2) & iOS     & B (Unique Scene) \\ \hline
9 & iPhone7Plus (2)  & iOS     & B (Unique Scene)  \\ \hline
10 & iPhoneX (2)      & iOS     & B (Unique Scene) \\ \hline
 
\end{tabular}
\end{center}
\end{table}

Cameraw operates much like any other camera application, using one button to capture a scene. When that button is pressed, Cameraw collects images with a variety of exposure settings, stepping through a pre-selected sequence of changing ISO and exposure times (EXP). From this sequence, we use the following three settings: (1) auto-exposed/nominal (camera establishes ISO and EXP automatically); (2) over-exposed/off-nominal (3 * ISO, 2 * EXP); and (3) under-exposed/off-nominal (0.5 * ISO, 0.5 * EXP). Camera aperture for each device remained constant (smartphone apertures cannot be changed).

\subsection{Images \& Data-Collection Protocols}
\label{ImAndDataProtocols}

\subsubsection{Images}
\label{Images}
The data comprise 8400 images from StegoAppDB \cite{NewmanLinChenReindersWangWuGuan2019}, with 300 images captured from each of 28 smartphones across ten brands/models (e.g., Apple/iPhone8); see Table \ref{DataDevices}. One third of all the images were auto-exposed; one third of all images were intentionally over-exposed; the remaining third were intentionally under-exposed. Full-sized images are used for the experiments described in this paper.

\subsubsection{Protocol}
\label{Protocol}
Table \ref{DataDevices} shows two protocols, A and B, that were followed during data-collection. One (A) is for same-scene content; the other (B) is for unique-scene content.

Protocol A: Same-scene content. Each of the 18 protocol-A cameras was attached to a tripod in a given scene location. Each camera took three images with staggered exposure settings, as previously described. All cameras were similarly oriented, so there were no upside-down images. This process was repeated 100 times, with 100 unique scene positions for the tripod. The scene content was the same for all 18 cameras, although the registration may have been slightly imperfect. 

Protocol B: Unique-scene content. Ten cameras follow a ``unique'' scene content acquisition protocol. This is the same as Protocol A except the scene content is not repeated from one camera to the next. In contrast to the Protocol-A images, which comprised 100 different scenes, Protocol-B images comprised 1000 different scenes.

\subsection{Exposure Value Comparison}
\label{EV}

One clear justification for our characterization of auto-, under-, and over-exposed images is the exposure value for the examples in Fig. \ref{ExposureSceneExamples}. Consider the three images of oranges in a bowl from the OnePlus5-1 camera (first column of the figure). The top-row image is visually much darker than the middle-row auto-exposed image, and the bottom-row image is visually much brighter. The under-exposed image has an ISO of 500, half of 1000 (the auto-exposed image ISO value), and an exposure time of 1/100 seconds (again, half of the auto-exposed image exposure time of 1/50 seconds). Similarly, the over-exposed image has an ISO of 3000, three times the auto-exposed ISO of 1000, and an exposure time of 1/25 seconds (twice the auto-exposed image exposure time of 1/50 seconds).

The exposure value is a quantification of light on the sensor determined by the $f$-stop (related to the aperture, which is constant in smartphone cameras), exposure time, and ISO. Exposure value is lower when there is less light and higher when there is more light on the sensor. We calculate exposure value as described in \cite{Lee2005} relative to the exposure value with ISO 100 ($EV_{100}$). For the auto-exposed example image, the exposure value is $EV_{100} + \log_2 10$. The under-exposed example image has an exposure value of $EV_{100} + \log_2 5$ and the over-exposed example has an exposure value of $EV_{100} + \log_2 30$. Clearly, the under-exposed image has quantifiably less and the over-exposed image has quantifiably more light on the sensor than the auto-exposed image. This relationship holds for all images in our dataset, which consists of only typically-lit indoor scenes.

\subsection{Visual Validation of Off-Nominal Settings}
\label{HumanValidation}

Another supporting argument for the three exposure types is a validation of the images using human judgment. By assessing the agreement amongst three human judges and one computer program, we can measure the consistency of the responses. If the consistency is high enough, we are satisfied that this validation supports using these images for testing PRNU camera identification of nominal and off-nominal exposures.

We used the same 5600 off-nominal images described in Section \ref{Images}, half of which are under-exposed and half are over-exposed. This is too many images for a human rater to evaluate in a reasonable amount of time without fatigue, so we randomly selected 5\% (280 images) for a human-judgment study. Half of those (140 images) were too dark and the other half were too light. Each half was mixed with 140 randomly drawn, nominally exposed images to form two sets of 280 images each.

The two sets of 280 images were shown to human judges in a web-based tool displaying a single image at a time. The judge clicked one of three text boxes indicating their judgment as auto/over/under-exposed. When a text box was clicked, the tool logged the choice, and advanced to the next image. The task took about 17 minutes per set.

The resulting data were analyzed using Fleiss’s kappa \cite{Fleiss1971} with four raters – three human judges and the computer program that chose the images in the first place. Using the ``R'' statistical software package \textit{irr} and the ``R'' function \textit{kappam.fleiss} \cite{GamerLemonFellowsSingh2012, RCoreTeam2017}, the kappa value was 0.9430212 with a z-statistic of 75.70786 (i.e., 75.7 standard deviations away from the mean) with a p-value of 0 – a nearly-perfect agreement amongst the four raters. This is clearly a level of confidence that justifies using these images for testing a camera-identification algorithm.


\section{Methods}
\label{Methods}

We propose a methodical investigation of off-nominally exposed images. The methodology consists of the following steps:\\

\begin{enumerate}
    \item Define a careful data collection protocol which minimizes the differences between nominal and off-nominal data, intentionally changing the characteristic under investigation.
    \item Isolate the points in the forensic algorithm where the characteristic under investigation impact the error rate estimate.
    \item Establish baseline error rates by executing the forensic algorithm using only the collected nominal data.
    \item For each point in the algorithm identified in step 2, incrementally and independently exchange the nominal data with the off-nominal data.
\end{enumerate}

We are investigating exposure settings, so our nominal dataset is our auto-exposed images and the off-nominal dataset consists of our over- and under-exposed images. Next, we identify estimation of the camera fingerprint (Box 1 in Fig. \ref{Standard PRNU Algorithm}) and the questioned image (Boxes 2 and 3 in Fig. \ref{Standard PRNU Algorithm}) as the two steps of the PRNU cameraID algorithm directly impacted by image exposure settings. Section \ref{MethodsAuto} details our baseline experiment, which generates camera fingerprints using auto-exposed images and uses a set of questioned images composed only of auto-exposed images. Finally, we iteratively change the exposure type of the images used to generate the camera fingerprint and the set of questioned images. This allows us to understand how off-nominal exposure images alter the error rate estimates for the PRNU cameraID algorithm, and provides baseline error rates for direct comparisons.

We investigate the impact of off-nominal exposure settings on the PRNU cameraID algorithm by partitioning the data into three exposure types. Each exposure type is used systematically to generate the camera fingerprints and/or questioned images. 

We performed 3 fundamental experiments: \\
1. Auto-exposed images (baseline / nominal) \\ 
2. Over-exposed images (too light / off-nominal) \\ 
3. Under-exposed images (too dark / off-nominal)

We also implement two sensitivity analyses, one for the sensitivity of the TPR to different proportions of off-nominal exposure images in the questioned image set and one for the sensitivity of error rates relative to the PCE decision threshold. A sensitivity analysis can show how even small, controlled changes impact error rate estimates.

\subsection{Auto-exposed images - baseline / nominal}
\label{MethodsAuto}

The baseline experiment is the PRNU cameraID algorithm (Section \ref{OverviewOfPRNUCameraIdentification}) applied to a set of nominally exposed images, which comprises 2800 auto-exposed images (100 auto-exposed images from each of the 28 cameras). We repeat the baseline experiment for five trials, where each trial is defined by a specific partitioning of the images.

First, we randomly partition the 100 images per camera into two groups: 1) 30 images used to generate the camera fingerprint; 2) 70 questioned images. The images are partitioned so that no camera fingerprint image shares scene content with any questioned image. For each camera, estimate the camera fingerprint (28 cameras $\times$ 1 camera fingerprint = 28 camera fingerprints). For each questioned image, estimate the image fingerprint (28 cameras $\times$ 70 images = 1960 image fingerprints).

Second, we calculate the PCE scores between each camera fingerprint and its own questioned images (28 camera fingerprints $\times$ 70 fingerprints for images from the camera = 1960 PCE scores). Next, we compare the PCE scores for images from the camera to the threshold of 60. If the PCE score is above 60, the image is a true positive and contributes to the TPR. If the PCE score is at most 60, the image is a false negative and contributes to the False Negative Rate (FNR = $1 -$ TPR).

Next, we calculate the PCE scores between each camera fingerprint and a set of questioned images from a different camera (28 specific-camera fingerprints $\times$ 70 fingerprints from images from a different camera $\times$ 27 different cameras for test = 52920 PCE scores). To calculate a balanced accuracy (i.e., an accuracy that responds equally to the TNR and TPR), we select a random subset of 1960 PCE scores from the 52920 PCEs for questioned images from a different camera. In order to avoid lucky or unlucky subsets of PCE scores, we perform the random selection of 1960 PCE scores 100 times. We compare the PCE scores for images from other cameras to the threshold of 60. If the PCE score is above 60, the image is a false positive and contributes to the FPR (FPR = $1 -$ TNR). If the PCE score is at most 60, the image is a true negative and contributes to the TNR. We calculate the accuracy for each of the 100 PCE score subsets using the 1960 PCE scores used to calculate the TPR, and the 1960 PCE scores used to calculate the TNR. Finally, we average the TPR, TNR, and accuracy values for the 100 PCE score subsets to calculate the error rates for each of the five trials.

The auto-exposed image baseline experiment is also subjected to two sensitivity analyses regarding the TPR and TNR: 1) for different proportions of off-nominal exposures in the questioned image set, and 2) for different PCE thresholds. For the first sensitivity analysis, we incrementally replace 1\% of the images comprising the questioned image set with off-nominal data, thus creating 101 questioned image sets (i.e., the first set has 100\% auto-exposed questioned images and 0\% off-nominally exposed images, the second set has 99\% auto-exposed questioned images and 1\% off-nominally exposed images, and so on with the final set comprising 100\% off-nominally exposed images). The sensitivity analysis of the PCE decision threshold is performed by incrementally shifting the threshold of 60 and recalculating the TPR and TNR values.

\subsection{Over-exposed images - too light / off-nominal}
\label{MethodsOver}

The over-exposed image experiment differs from the baseline by estimating two camera fingerprints (one using auto-exposed images and one using over-exposed images) and the questioned image set comprises only over-exposed images. This experiment uses 5600 images: 100 auto-exposed images from each of the 28 devices and 100 over-exposed images from each of the 28 devices.

We adapt the procedure used in the baseline experiment for the over-exposed experiment by performing the same procedural steps but with different sets of data: 1) 30 auto-exposed images per camera are used to estimate the camera fingerprints and 70 over-exposed images from each camera comprise the set of questioned images (auto-fingerprint vs. over-image); 2) 30 over-exposed images per camera are used to estimate the camera fingerprints and 70 over-exposed images from each camera comprise the set of questioned images (over-fingerprint vs. over-image). We repeat the sensitivity analysis of the PCE threshold on the auto-fingerprint vs. over-image and over-fingerprint vs. over-image experiments.

We remark that the most likely scenario where a forensic practitioner might encounter an over-exposed image is as a questioned image. The practitioner may have access to the suspect camera and can therefore control the exposure settings of images used to estimate the camera fingerprint, but the exposure of the questioned image is already established. A forensic practitioner using over-exposed images for the camera fingerprint when the questioned image is auto-exposed is an unlikely scenario, so we omit the results of this experiment. We investigate the over-fingerprint vs. over-image scenario as a possible solution to the degraded error rates caused by comparing auto-exposed fingerprints and over-exposed questioned images. This experiment examines whether camera fingerprints estimated with the same exposure type as the questioned image will have a more similar PRNU estimate than camera fingerprints estimated solely from auto-exposed images, potentially reducing the error rates.

The sensitivity analysis of the PCE decision threshold is performed for both experiments by incrementally shifting the PCE threshold and recalculating the TPR and TNR values.

\subsection{Under-exposed images - too dark / off-nominal}

The under-exposed image experiment repeats the investigations in Section \ref{MethodsOver} except we replace the over-exposed images with under-exposed images, as characterized in Section \ref{Data} — Data. The purpose behind each of the two under-exposed image scenarios corresponds to the motivations for the over-exposed experiment scenarios. Each investigation was also subjected to a sensitivity analysis of the PCE threshold.

\section{Results}
\label{Results}

We present results from the experiments detailed in Section \ref{Methods} - Methods for images of each exposure type. The TPR decreases by at least 14.27\% for the off-nominal questioned images compared to the auto-exposed questioned-image baseline, meaning many images from the specific camera are missed. Similarly, the TNR decreases to 99.46\% for under-exposed questioned images - an error of approximately one in two hundred images incorrectly identified as a match with the specific camera. Such mistakes can lead to increased false positives, often connected to wrongful convictions. Our results are separated into nominal and off-nominal questioned images, and end with an investigation of possible remediations (Sections \ref{NominalResults}, \ref{OffNominalResults}, and \ref{Solutions}, respectively).

\begin{table}[!h]
\centering
\caption{Aggregate results over 5 trials averaged over 100 repetitions}
\label{BalancedAggregateResults}
\begin{tabular}{|l|c|c|c|}
\hline
\multicolumn{1}{|c|}{\textbf{Experiment}} & \textbf{TPR (std.)} & \textbf{TNR (std.)} & \textbf{Accuracy (std.)} \\ \hline 
\begin{tabular}[c]{@{}l@{}} {\it Auto-fingerprint} \\ {\it vs. Auto-image Baseline}\end{tabular} & \makecell{{\it 1} \\ {\it (0)}} & \makecell{ {\it 0.9992} \\ {\it (0.0001)}} & \makecell{ {\it 0.9996} \\ {\it (0.0001)}} \\ \hline \hline
\begin{tabular}[c]{@{}l@{}} Auto-fingerprint \\ vs. Over-image\end{tabular} & \makecell{0.8290 \\ (0.0030)} & \makecell{0.9998 \\ (0.0001)} & \makecell{0.9144 \\ (0.0015)} \\ \hline
\begin{tabular}[c]{@{}l@{}} Auto-fingerprint \\ vs. Under-image\end{tabular} & \makecell{0.8573 \\ (0.0003)} & \makecell{0.9946 \\ (0.0007)} & \makecell{0.9260 \\ (0.0004)} \\ \hline \hline
\begin{tabular}[c]{@{}l@{}}Over-fingerprint \\ vs. Over-image\end{tabular} & \makecell{0.8888 \\ (0.0070)} & \makecell{0.9996 \\ (0.0002)} & \makecell{0.9442 \\ (0.0035)} \\ \hline
\begin{tabular}[c]{@{}l@{}} Under-fingerprint \\ vs. Under-image \end{tabular} & \makecell{0.9999 \\ (0.0002)} & \makecell{0.8426 \\ (0.0043)} & \makecell{0.9213 \\ (0.0020)} \\ \hline
\end{tabular}
\end{table}


\begin{table*}[!hbt]
\centering
\caption{True Positive Rate (TPR), True Negative Rate (TNR), and standard deviation (STD) for auto-exposed camera fingerprint with auto-exposed questioned images (left column), over-exposed test data (middle column), and under-exposed test data (right column) by camera model. TPRs and TNRs lower than the baseline experiment are in {\bf BOLD} for the off-nominal experiments.}
\label{ModelResults}
\begin{tabular}{|l|cc|cc|cc|}
\hline
\textbf{Camera Model} & \multicolumn{2}{c|}{\textit{\textbf{Auto-fingerprint vs. Auto-image Baseline}}} & \multicolumn{2}{c|}{\textbf{Auto-fingerprint vs. Over-image}} & \multicolumn{2}{c|}{\textbf{Auto-fingerprint vs. Under-image}} \\ \hline

 & \multicolumn{1}{c|}{\textit{\textbf{TPR (STD)}}} & \textit{\textbf{TNR (STD)}} & \multicolumn{1}{c|}{\textbf{TPR (STD)}} & \textbf{TNR (STD)} & \multicolumn{1}{c|}{\textbf{TPR (STD)}} & \textbf{TNR (STD)} \\ \hline
\textbf{Pixel1} & \multicolumn{1}{c|}{\textit{1 (0)}} & \textit{0.9998 (0.0002)} & \multicolumn{1}{c|}{\textbf{0.9850 (0.0122)}} & \textbf{0.9995 (0.0003)} & \multicolumn{1}{c|}{1 (0)} & \textbf{0.9857 (0.0021)} \\ \hline
\textbf{Pixel2} & \multicolumn{1}{c|}{\textit{1 (0)}} & \textit{0.9998 (0.0003)} & \multicolumn{1}{c|}{\textbf{0.0007 (0.0016)}} & 0.9999 (0.0002) & \multicolumn{1}{c|}{\textbf{0.0014 (0.0020)}} & \textbf{0.9977 (0.0010)} \\ \hline
\textbf{iPhone6s} & \multicolumn{1}{c|}{\textit{1 (0)}} & \textit{0.9999 (0.0001)} & \multicolumn{1}{c|}{\textbf{0.9936 (0.0016)}} & \textbf{0.9997 (0.0001)} & \multicolumn{1}{c|}{1 (0)} & \textbf{0.9984 (0.00003)} \\ \hline
\textbf{iPhone7} & \multicolumn{1}{c|}{\textit{1 (0)}} & \textit{0.9999 (0.0001)} & \multicolumn{1}{c|}{\textbf{0.9750 (0.0067)}} & 0.9999 (0.0001) & \multicolumn{1}{c|}{1 (0)} & \textbf{0.9995 (0.0003)} \\ \hline
\textbf{iPhone8} & \multicolumn{1}{c|}{\textit{1 (0)}} & \textit{1 (0)} & \multicolumn{1}{c|}{\textbf{0.9543 (0.0130)}} & \textbf{1.0000 (0.0001)} & \multicolumn{1}{c|}{1 (0)} & \textbf{0.9998 (0.0003)} \\ \hline
\textbf{OnePlus5} & \multicolumn{1}{c|}{\textit{1 (0)}} & \textit{0.9984 (0.0009)} & \multicolumn{1}{c|}{\textbf{0.9957 (0.0039)}} & 0.9993 (0.0006) & \multicolumn{1}{c|}{1 (0)} & \textbf{0.9673 (0.042)} \\ \hline
\textbf{SamsungS8} & \multicolumn{1}{c|}{\textit{1 (0)}} & \textit{0.9920 (0.0018)} & \multicolumn{1}{c|}{1 (0)} & 0.9999 (0.0001) & \multicolumn{1}{c|}{1 (0)} & 0.9957 (0.0005) \\ \hline
\textbf{iPhone6sPlus} & \multicolumn{1}{c|}{\textit{1 (0)}} & \textit{1 (0)} & \multicolumn{1}{c|}{\textbf{0.9914 (0.0032)}} & \textbf{0.9999 (0.0002)} & \multicolumn{1}{c|}{1 (0)} & \textbf{0.9992 (0.0002)} \\ \hline
\textbf{iPhone7Plus} & \multicolumn{1}{c|}{\textit{1 (0)}} & \textit{1 (0)} & \multicolumn{1}{c|}{\textbf{0.8729 (0.0155)}} & \textbf{1.0000 (0.0001)} & \multicolumn{1}{c|}{1 (0)} & \textbf{0.9994 (0.0004)} \\ \hline
\textbf{iPhoneX} & \multicolumn{1}{c|}{\textit{1 (0)}} & \textit{1 (0)} & \multicolumn{1}{c|}{\textbf{0.8829 (0.0139)}} & \textbf{1.0000 (0.0001)} & \multicolumn{1}{c|}{1 (0)} & \textbf{0.9999 (0.0002)} \\ \hline
\end{tabular}
\end{table*}

\subsection{Baseline (Nominal Images)}
\label{NominalResults}

The case where the camera fingerprint and questioned images are auto-exposed provides a baseline to compare with the off-nominal experiments. Results from this baseline experiment are shown in the first row of Table \ref{BalancedAggregateResults}, and represent the current scenario used by forensic practitioners. The TPR is 100\% and the TNR is 99.92\%. Results for the same baseline data are listed by the 10 individual models, shown on the left-hand side of Table \ref{ModelResults}. These results are in line with the error rates from the study in \cite{GoljanFridrichFiller2009}.

\subsection{Off-Nominal Questioned Image Sets}
\label{OffNominalResults}

The most singular results using off-nominal data with the PRNU cameraID algorithm are the TPR values shown in Table \ref{BalancedAggregateResults}. Rows two and three in Table \ref{BalancedAggregateResults} are error rate estimates for the most common scenarios where a practitioner might encounter an off-nominally exposed image: when the camera fingerprint is estimated from auto-exposed images, but the questioned image is over- or under-exposed. The TPR values for these two experiments are strikingly different from the baseline results. When the camera fingerprint is composed of auto-exposed images and all questioned images are over-exposed, the TPR is 82.90\% and the TNR is 99.98\% (row two, Table \ref{BalancedAggregateResults}). When the camera fingerprint is composed of auto-exposed images and all questioned images are under-exposed, the TPR is 85.73\% and the TNR is 99.46\% (row three, Table \ref{BalancedAggregateResults}). The TPRs for these off-nominal images – a TPR decrease of 17.1\% for over-exposed and of 14.27\% for under-exposed - are much lower than the 100\% TPR for the baseline experiment. The large reduction in TPR from our baseline experiment warrants attention to the differing error rates between image exposure types. Additionally, the TNR of 99.46\% for the under-exposed questioned images (row three, Table \ref{BalancedAggregateResults}) corresponds to an FPR of 0.54\%. This is an error of roughly one in two hundred associated with incorrectly matching an image with a camera.

\begin{figure}[h!]

\centerline{\includegraphics[scale = 1]{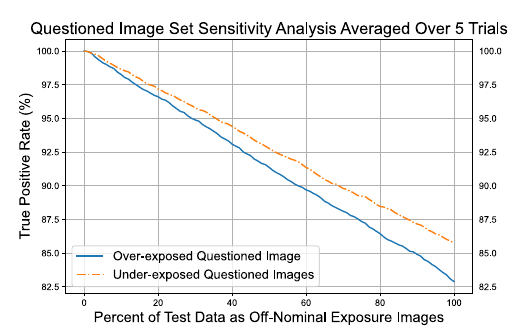}}
\caption{\textbf{Sensitivity Analysis of Off-Nominally Exposed Questioned Images:} Both under-exposed (dashed orange line) and over-exposed (solid blue line) questioned images have a roughly linear relationship with the degradation of the True Positive Rate (TPR). This implies that even small percentages of off-nominally exposed images in the questioned image set can have an impact on the error rate estimates.}
\label{QuestionedImageSetSensitivityAnalysis}
\end{figure}

The sensitivity analysis of the TPR when the questioned image set consists of various proportions of off-nominal images is given in Fig. \ref{QuestionedImageSetSensitivityAnalysis}. This sensitivity analysis highlights the negative linear relationship between the proportion of off-nominally exposed questioned images and the TPR. Both the over-exposed questioned images (Fig. \ref{QuestionedImageSetSensitivityAnalysis}, solid blue line) and the under-exposed questioned images (Fig. \ref{QuestionedImageSetSensitivityAnalysis}, dashed orange line) have a direct impact on the TPR estimate, even when only a small percent of the questioned image set are off-nominally exposed. The leftmost endpoints of the orange and blue lines in Fig. \ref{QuestionedImageSetSensitivityAnalysis} correspond to TPRs of 100\% (TPR for row one of Table \ref{BalancedAggregateResults}). Similarly, the rightmost endpoints of the orange (dashed) and blue (solid) lines in Fig. 3 correspond to a TPR of 82.90\% for over-exposed images (row two of Table \ref{BalancedAggregateResults}) and 85.73\% for under-exposed images (row three of Table \ref{BalancedAggregateResults}). Note that the TPR for over-exposed images is always less than the TPR for under-exposed images, regardless of the proportion of off-nominal test data. The consistency of this linear relationship supports the notion that off-nominal exposure types do indeed affect the error rates of the PRNU cameraID algorithm.

The results for the auto-exposed camera fingerprint experiments are listed for the 10 models in Table \ref{ModelResults} (auto-image in the left-hand column, over-image in the middle column, and under-image in the right-hand column). Note that the Pixel 2 performs dramatically poorly for both off-nominal exposures with a TPR of 0.07\% on over-exposed questioned images and TPR of 0.14\% on under-exposed questioned images. Although determining the cause of the Pixel 2 camera model's poor performance for off-nominally exposed images is outside the scope of this paper, we theorize this could be caused by proprietary pipeline processing, which is protected by manufacturers (in this case, Google). Poor performance for specific camera models has been observed in prior research, such as the 0.8\% TNR of the Fujifilm X-T30 in \cite{IulianiFontaniPiva2021}. Particularly poor performance by individual models is another reason to encourage rigorous tool validation procedures \cite{HughesKarabiyik2020}.

\subsection{Possible Solutions for Off-Nominal Data}
\label{Solutions}

\begin{figure*}[!h]

\centerline{\includegraphics[scale = 1]{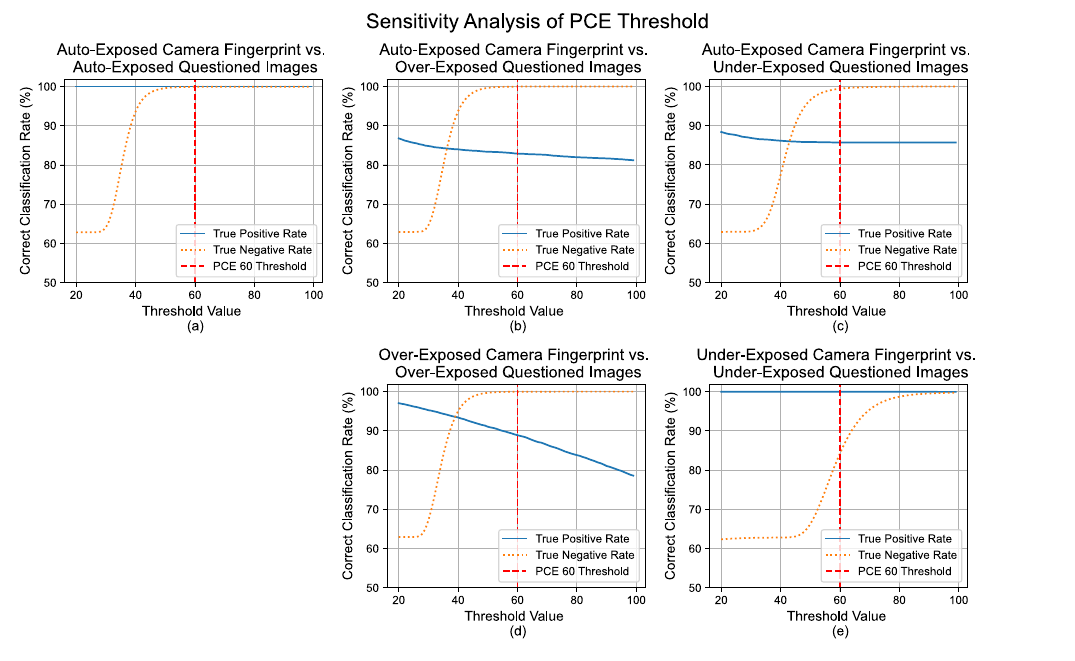}}
\caption{\textbf{Sensitivity analysis of PCE threshold (off-nominal exposures).} The upper left graph (baseline) shows a near-perfect TPR and TNR for the PCE threshold of 60 (dashed red line) when both the camera fingerprint and the image fingerprint are estimated from nominal images. However, when fingerprints are estimated from any off-nominal images, either the TPR or the TNR degrades dramatically. Raising the PCE threshold does not improve the TPR (solid blue line), although it does gradually improve the TNR (dotted orange line). Lowering the PCE threshold slightly improves the TPR, but does degrade the TNR.}
\label{ThresholdSensitivityAnalysisUE}
\end{figure*}

One possible method to improve the error rates for off-nominal exposures is to estimate the camera fingerprint from the same exposure type as the questioned image. When the camera fingerprint uses only over-exposed images and the questioned images are also over-exposed (row 4 in Table \ref{BalancedAggregateResults}), the TPR rises slightly to 88.88\%, but the TNR falls to 99.96\%. Similarly, when the camera fingerprint uses only under-exposed images and the questioned images are also under-exposed (row 5 in Table \ref{BalancedAggregateResults}), the TPR increases considerably (99.99\%), yet the TNR decreases (84.26\%). A drop in the TNR corresponds to an increase in false positives, so this is a trade-off that forensic practitioners are unlikely to accept.

Another possible solution to the reduced accuracy values in Table \ref{BalancedAggregateResults} for off-nominal exposures is to use an alternative PCE threshold. We perform a sensitivity analysis of the TPR and TNR to different PCE thresholds by incrementally shifting the integer PCE threshold and recalculating the TPR and TNR using the same set of PCE scores. Fig. \ref{ThresholdSensitivityAnalysisUE} demonstrates that lowering the threshold to increase the TPR for the over-exposed image experiments (plots (b) and (d) in Fig. \ref{ThresholdSensitivityAnalysisUE}) or for the under-exposed image experiments (plots (c) and (e) in Fig. \ref{ThresholdSensitivityAnalysisUE}) would also lower the TNR, meaning that the overall accuracy would not improve.

Our final experiment also investigates an alternate PCE threshold. The PCE threshold of 60 was initially set to produce a 0\% FPR on the subset of images from a different model than the specific camera \cite{GoljanFridrichFiller2009}. We compute the lowest integer threshold value to produce a 0\% FPR (regardless of camera model) and recalculate the accuracy (results for each experiment are given in Table \ref{ThresholdSAResults}). Note that in each case the threshold must be raised, and in some cases more than doubled (rows 1 and 3 through 5 in Table \ref{ThresholdSAResults}). For the over-exposed image experiments, these accuracy values are all lower than the accuracy using the PCE threshold of 60 (rows 2 and 4 in Table \ref{ThresholdSAResults}). The alternate threshold does improve the accuracy slightly for the auto-fingerprint vs. under-image experiment (row 3 in Table \ref{ThresholdSAResults}) compared with the results for the PCE threshold of 60 (row 3 in Table \ref{BalancedAggregateResults}). Choosing a threshold based on a 0\% FPR, however, not only depends on the dataset and exposure type, but is problematic for other reasons detailed in Section \ref{Discussion}.

\begin{table}[!h]
\centering
\caption{Lowest threshold resulting in a 100\% TNR (0\% FPR) for each experiment and the corresponding TPR and accuracy.}
\label{ThresholdSAResults}
\begin{tabular}{|l|c|c|c|c|}
\hline
\multicolumn{1}{|c|}{\textbf{Experiment}} & \textbf{Threshold} & \textbf{TPR} & \textbf{TNR}  & \textbf{Accuracy} \\ \hline
\begin{tabular}[c]{@{}l@{}} {\it Auto-fingerprint} \\ {\it vs. Auto-image Baseline} \end{tabular}   & {\it 210}     & {\it 1}            & {\it 1}      & {\it 1}      \\ \hline \hline
\begin{tabular}[c]{@{}l@{}} Auto-fingerprint \\ vs. Over-image\end{tabular}  & 77   & 0.8213       & 1   & 0.9107         \\ \hline
\begin{tabular}[c]{@{}l@{}} Auto-fingerprint \\ vs. Under-image\end{tabular}   & 143  & 0.8569       & 1  & 0.9285          \\ \hline \hline
\begin{tabular}[c]{@{}l@{}} Over-fingerprint \\ vs. Over-image\end{tabular}  & 442   & 0.2300       & 1    & 0.6150        \\ \hline
\begin{tabular}[c]{@{}l@{}} Under-fingerprint \\ vs. Under-image\end{tabular}   & 539  & 0.7587       & 1 & 0.8794           \\ \hline
\end{tabular}
\end{table}

\section{Limitations}
\label{Limitations}

Digital image forensics researchers typically encounter one of two problems for data collection: resource exhaustion or quality of images. Manually taking enough pictures from a large variety of cameras is often infeasible due to the significant time and resources required. An alternative source of data collection is scraping images from online collections (e.g., Flickr). However, these images are of unknown provenance and their true origin and exposure settings are frequently unknown.

The foundational PRNU cameraID research \cite{GoljanFridrichFiller2009} prioritized the size of the dataset by downloading over one million Flickr images from nearly seven thousand cameras, enabling them to make universal claims. The absence of a data-collection protocol makes it impossible to ascertain the proportion of nominal/off-nominal images in that dataset. We chose instead to prioritize control over the image collection and exposure settings by taking 8400 calibrated images from 28 cameras, as detailed in Section \ref{Data} - Data. Although we are not claiming universality, we present our results as a demonstration that off-nominal exposure settings can alter camera-identification error rates, often quite dramatically. Our claims about our dataset are only possible because of the exacting data-collection protocol followed. A similar, but larger-scale study remains for future research.

A further limitation is that the cameras used in our research are not the most recent models (e.g., the iPhone X was released in November 2017). We have estimated that it would take at least 14 weeks to add just one more camera of a new model. Adding a new camera would delay the preparation of a technical report by roughly a quarter of a year, by which time yet another new model would likely have been released. Therefore, it is simply not practical to continually add up-to-date cameras for the purpose of one paper. Newer camera models include additional complications that must be addressed in future work, including high dynamic ranges, multiple lenses and sensors, new proprietary processing, AI, etc. That said, we recognize the importance of maintaining current datasets if camera-identification technology is to keep pace with camera development.

\section{Discussion}
\label{Discussion}

The off-nominal image experiments performed dramatically poorly when compared with the auto-exposed image baseline. The 14-17\% TPR decrease is quite large (rows two and three, Table \ref{BalancedAggregateResults}) and cannot be attributed to chance, poor data quality, or methodological errors. When used to investigate which camera captured an evidence image, a TPR of 85.73\% is roughly one out of seven, meaning the correct camera might be missed as many as one out of seven times if the questioned image is under-exposed (or more often if the questioned image is over-exposed).

One goal of forensic research is to prepare methodologies for use in a court of law. The PRNU camera-identification algorithm described in \cite{GoljanFridrichFiller2009} forms the foundation of an FBI application called FINDCamera \cite{MacKenzieBruehs2018}. The false-positive error rate estimate of one in a million presented by an expert witness was based on over one million images from Flickr \cite{USDistrictCourt2011b}. Our results show that the false-positive error rate can rise to one in two hundred for under-exposed questioned images, and one in five thousand for over-exposed questioned images. Differences in error rate estimates could impact how jurors weigh the strength of evidence, which is particularly important in cases with severe consequences. For example, the aforementioned 2011 trial resulted in a prison sentence of 45 years \cite{USDistrictCourt2011b}.

We investigated two seemingly-obvious solution candidates that, unfortunately, turned out not to fully address the aforementioned shift in error rates. The first solution was to create camera fingerprints consistent with the exposure type of the questioned image. However, this approach only modestly improves the over-exposure true-positive error rates, while causing the false-positive rate to skyrocket for under-exposed images (rows 4 and 5 of Table \ref{BalancedAggregateResults}). A second solution is to use a PCE threshold other than 60, as demonstrated in Fig. \ref{ThresholdSensitivityAnalysisUE} and in Table \ref{ThresholdSAResults}. Note that the alternate thresholds only minimally improve the error rates, and in some cases simply exacerbate the problem.  Additionally, recent work has raised concerns that examiners changing decision thresholds based on subjective analysis of the evidence and forensic scenario can negatively impact the legal system \cite{Thompson2023}. 

Although there is a good instinct to mitigate any error inherent to a methodology, it can be difficult to change existing forensic tools. Further refining error rates in context of the questioned image can help forensic practitioners better understand and communicate the error associated with pre-existing tools. Introducing new methodologies requires both acceptance by forensic practitioners and rigorous research to meet the Daubert standard, which requires time, use by others in the community, and passing another Daubert challenge for the new tool. Methods which can be used in conjunction with current tools (such as our proposed protocol to estimate more accurate error rates with respect to the exposure type of the questioned image) can introduce needed incremental changes between significant technological shifts.

Our experiments clearly demonstrate that off-nominal images (e.g., over- or under-exposed) impact the error rates of the PRNU cameraID algorithm. We also show that the two most obvious and straightforward modifications to the existing algorithm do not adequately rectify the performance problems for off-nominal exposure images. Proper tool validation and error rate estimation is a crucial aspect of this forensic field that must continually be improved.


\section{Conclusions}
\label{Conclusions}

We present a study of the PRNU source-camera-identification algorithm \cite{GoljanFridrichFiller2009} for off-nominal (over- and under-exposed) images using a meticulously-collected dataset \cite{NewmanLinChenReindersWangWuGuan2019}. Our work implements a systematic investigation to show that error rates are worse when off-nominal images are used for forensic source camera identification. In particular, for over-exposed questioned images the true-positive rate is 82.90\%, as compared with 100\% for nominal (auto-exposed) images. Of note is the contrast between our nominal baseline's false-positive rate (0.08\%) and the roughly one in two hundred false-positive rate for under-exposed (too dark) images. This disparity is concerning, as it can have real-life consequences in the criminal judicial process. Simple and obvious mitigations, such as changing PCE thresholds, do not solve the error-rate problem for off-nominal images. The insight gained from our methodology can help forensic practitioners better understand and communicate the error rates of forensic tools when applied to data representing off-nominal conditions.


\appendices
\section{Data and Code Availability}

Data from this study are available upon request. Code is available from its authors \cite{GoljanChenComeanaFridrichMATLAB}.


\section{Attributions/Declarations}

\begin{flushleft}
   Author contributions: 
\end{flushleft}

AM: Conceptualization, analysis, initial/final drafts.

RM: Conceptualization, methodology, analysis, all drafts. 

JN: Conceptualization, funding acquisition, initial draft.

\begin{flushleft}
    Ethical approval:
\end{flushleft}

\begin{adjustwidth}{1em}{}
The Iowa State University Institutional Review Board confirmed that the human-subjects aspect of this study is exempt; the information collected contains no personally identifiable information, and is not intended to contribute to generalizable knowledge.
\end{adjustwidth}

\begin{flushleft}
   Conflicts: 
\end{flushleft}

The authors declare no conflicts of interest.


\section*{Acknowledgment}

We thank Huayun Huang for her design and implementation of the user-study data-set validation software.


\bibliographystyle{ieeetr}
\bibliography{master_bib}


\begin{IEEEbiography}[{\includegraphics[width=1in,height=1.25in,clip,keepaspectratio]{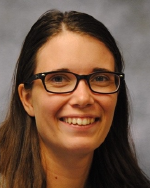}}]{Abby Martin} received her B.A. degree in Computer Science/Software Engineering and Mathematics from Augustana University, Sioux Falls, SD, in 2017 and is pursuing her Ph.D. degree in Mathematics and Computer Science from Iowa State University, Ames, IA under the supervision of Professors Jennifer Newman and Jin Tian. From 2020 to 2024, she was a Research Assistant with the Center for Statistics and Applications in Forensic Evidence. Her research interests include camera source identification, steganalysis, and digital image forensics.
\end{IEEEbiography}

\begin{IEEEbiography}
{Roy Maxion} (IEEE Fellow) is a Research Professor emeritus in computer science and machine learning at Carnegie Mellon University, where he is also the director of the Dependable Systems Laboratory. He has long been a passionate proponent of rigorous/foundational scientific methodology. He is an IEEE Fellow, and recently served as a member of the US National Academy of Sciences committee on Future Research Goals and Directions for Foundational Science in Cybersecurity. He won the 2019 IEEE/IFIP Test-of-Time award for his work using typing rhythms for user authentication.  He recently won the Taiwan Tamkang Panda Trophy for his lecture on the sensitivity of machine learning systems to small irregularities in data. He is one of the founding members of the US Center for Statistics and Applications in Forensic Evidence.  He is a member of the editorial boards of IEEE Security and Privacy and the International Journal of Machine Learning.
\end{IEEEbiography}

\begin{IEEEbiography}[{\includegraphics[width=1in,height=1.25in,clip,keepaspectratio]{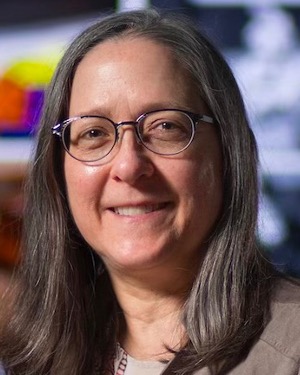}}]{Jennifer Newman} Jennifer L. Newman (Senior Member, IEEE) is a Professor of Mathematics at Iowa State University, Ames, Iowa and holds a Scott Hanna Faculty Fellow in Mathematics. She received her B.A. in Physics from Mount Holyoke College, and an M.S. and Ph.D. in Mathematics from the University of Florida, Gainesville. She has served as an Associate Editor for the Journal of Electronic Imaging and for the Journal of Mathematical Imaging and has served on numerous program committees for many professional conferences. Her current research is in statistical forensic imaging, including steganography, steganalysis and camera identification. Other research interests include image algebra; object detection, modeling and machine learning for images; morphological and other neural networks; and statistical modeling of textures for forward and inverse problems. She has over 80 refereed publications and has been a Principal Investigator or Co-Principal Investigator on over \$24 million in grants. She has mentored over 32 graduate students as their major professor, and many undergraduate students as well.
\end{IEEEbiography}

\EOD

\end{document}